# PHASE-RESOLVED SPECTROSCOPY OF GEMINGA SHOWS ROTATING HOT SPOT(S)

P.A. Caraveo [1], A. De Luca[1], S. Mereghetti[1], A. Pellizzoni [1] and G.F. Bignami [2,3,1]

1) IASF-CNR, Via Bassini, 15- 20133 Milano, Italy

2) CESR/CNRS-UPS, 9 Av. du Colonel Roche, Toulouse, France

3) Università di Pavia, Dip. Fisica Teorica e Nucleare, Via U. Bassi, 6 Pavia, Italy

**Isolated neutron stars (INS) are seen in x-rays through their non-thermal and/or surface thermal emissions. XMM-Newton observations of the Geminga pulsar show a 43 eV spectrum from the whole neutron star surface plus a power-law component above 2 keV. In addition, we have detected a hot (170 eV) thermal emission from a ~60 m radius spot on the pulsar's surface. Such a thermal emission, only visible at selected phase intervals, may be coming from polar hot spot(s), long thought to exist as a result of heating from magnetospheric accelerated particles. It may provide the missing link between the X and gamma ray emission of the pulsar.**

Photons emitted by pulsars carry the signature of their production mechanisms as well as of the geometry of their emitting regions. While neutron star physics is reflected in their photon spectra, geometrical constraints, such as viewing angles of rotational and magnetic axes, shape their observed light curve. Phase modulation takes place as different emitting regions are brought into view during the star rotation. Geometry can also influence source spectral shapes, because of different emission mechanisms in different regions.

In spite of the potential interest of phase-resolved spectroscopy, the paucity of detected x-ray photons made it so far impossible to apply this method to INS, but for the Crab pulsar[1]. EPIC on XMM-Newton and Chandra can now provide an adequate harvest of time-tagged photons. However, phase-resolved spectroscopy is not yet commonly used : so far it has been applied only to the Crab pulsar with Chandra[2], and to 1E1207-5209 with EPIC [3,4]. Although interesting, these sources represent specific and somewhat extreme cases amongst x-ray emitting neutron stars. The Geminga pulsar, on the other hand, is often considered as archetypal[5] for middle-aged (350.000 y) neutron stars, which emit x-rays mostly, but not only, owing to their surface thermal emission. INS surface temperatures frequently yield radiation in the X-ray domain[6], but keV photons can also be produced by energetic electrons in their strong magnetic fields. Geminga has the interesting characteristics of showing both thermal[7] and non-thermal processes to be at work in the sub-keV to several keV range[8,9]. With a photon number more than doubling all previous statistics, and with a wider (0.15 to 8 keV) spectral range, EPIC offers now the first chance of a meaningful phase-resolved spectroscopy for Geminga.

XMM-Newton performed a 100 ksec exposure on 4 April, 2002 using its three EPIC cameras. The two MOS [10] operated in "full frame" mode, while the pn one [11] operated in "small window" mode, ideal for accurate timing of source photons. After removing intervals with high particle background and correcting for dead time, we obtain a net exposure of 55.0 ksec for the pn camera and of 76.9 and 77.4 ksec for MOS1 and MOS2 respectively. The EPIC observation yielded a total of 76,850 photons in the energy range $0.15<E<8$ keV , the majority of which (52,850 photons) are due to the pn detector. The MOS images have unveiled two tails of diffuse emission trailing Geminga and well aligned with the source proper motion[12]. Here, we only present the analysis of the pn data, which were processed with the XMM Newton Science Analysis Software (SAS version 5.4.1).

First, we have fitted the time-averaged, total source data (Fig. 1) using a combination of a black body plus a power-law. In view of the unsatisfactory result, we added a third component, both in the form of a black-body and of a power-law. The resulting $\chi^2$ did improve significantly, suggesting that Geminga's spectrum indeed requires a three-component model. Since an additional, steep power-law cannot easily meet the constraints of the optical-to-UV part of the Geminga spectrum[13], we choose a second, hotter black-body component for the description of our time-averaged spectrum. To make our results comparable to the previous models [7,8,9], we adopted the $N_H$ value obtained by ROSAT[8] ($N_H = 1.07\ 10^{20}$). At the parallactic distance of 160 pc[14], the resulting parameters ($T_1 = 43 \pm 1$ eV ($4.8\ 10^5$ °K) from a $8.6 \pm 1$ km radius surface, $T_2 = 170 \pm 30$ eV ($1.9\ 10^6$ °K) from a $40 \pm 10$ m radius surface, power law photon index $1.7 \pm 0.1$) are similar to those derived by Halpern and Ruderman [7], and the index of the power-law is identical to the latest model from ASCA data[9]. We note that, when extrapolated to E> 100 MeV, such a power-law is two orders of magnitude below the detected gamma-ray flux [15].

We searched next for the best timing parameters. Standard epoch-folding for our data set yielded a "best" period of $0.2371012 \pm 1\ 10^{-7}$ sec. This agrees with the extrapolation of the EGRET ephemeris[9], which gives a period of 0.2371012153 sec, more accurate than our best value owing to the much longer time span of the γ-ray data. To compare our light curves with the EGRET ones, we have adopted the extrapolated EGRET period. Light curves for three energy intervals, i.e. 0.15-0.7 keV (dominated by cool black-body emission), 0.7-2 keV (hot black-body) and 2-8 keV (power-law), were drawn (Fig. 2).

Geminga's light curves are different for different energies[9]. The broad maximum at low energies is seen to change into two peaks at higher energies (0.7-2 keV and 2-8 keV): a prominent one at phase ~ 0.5 and a smaller peak at ~0.9. The minimum at phases 0.1-0.3, on the other hand, is common to all light curves. The pulsed fraction varies significantly as a function of energy, ranging from $30.3 \pm .7$ % at low energy to $54.5 \pm 2.4$ % in the middle range and back to $33.2 \pm 4.5$ % at higher energy. Our improved statistics does not confirm the very high pulsed fraction seen by[9] above 2 keV.

The γ-ray peaks occur at absolute pulsar phases 0.55 and 1.05, trailing our x-ray peaks by about 0.05- 0.1 of phase. However, such a phase comparison[9] is entirely dominated by the uncertainty on the source $\dot{P}$. Folding the EPIC data with $\dot{P}$ values within ± 2σ of the best Egret fit, we obtain phases varying by about ± 0.15. Such an error could shift our two X-ray peaks, separated by about 0.4 in phase, exactly between the two gamma-ray ones. However, a shift in the opposite direction would be equally probable.

The analysis done so far confirms and refines the results of two decades of x-ray studies of Geminga, from the Einstein satellite[16] to the sequence of ROSAT and ASCA data[7,8,9]. Spread among several ROSAT and ASCA observations, the pre EPIC X-ray photons from Geminga number ~33,000. With ~53,000 photons from a single observation, it makes sense to now try phase-resolved spectroscopy.

Accordingly, we have divided the pulsar light curve into 10 phase intervals, also shown in Fig. 2, and for each of them a separate spectrum was drawn (Fig.3A). Significant spectral changes as a function of phase are apparent (Fig. 3A and videoclip). To better assess such a phase-dependent behavior, we also considered the deviations of each individual spectrum from the phase-averaged one (Fig.S1 and S2). The biggest deviations occur in the medium interval (0.7 < E < 2 keV), dominated by the hot black-body. While spectra collected in intervals #1, 2 and 3 lack medium energy photons, spectra # 5,6 and 7 (the X-ray prominent peak) show an overabundance of such photons, consistent with previous observations of a slightly softer main peak [9].

Individual spectra were then fitted using the same three components which describe the average spectrum, with $N_H$ fixed to the ROSAT value[8].

First, the temperatures of both black bodies as well as the power-law index were allowed to vary. The resulting best fits showed that, within statistics, neither the two temperature values nor the power-law photon index had significant variations. Thus, we repeated the fitting exercise keeping temperatures and power-law index fixed, while leaving their normalizations as free parameters. Fig3B (with videoclip) gives the result of such a spectral fitting. For each phase interval, the three spectral components are shown. The phase-resolved spectra show that we need a second, hotter black-body, since its contribution, although strongly phase dependent, is unavoidable for a satisfactory fit to all spectra but #2, for which the hot black-body component is seen to vanish. This particular phase interval is well fit with just a single cool black-body and a power-law. The values derived for the emitting surfaces of the two black-bodies and for the power-law normalization vary as a function of the pulsar phase, albeit with different time evolutions (Fig. 4). While the variations of the emitting area of the two black bodies could be described as sinusoidal, the variation of the non-thermal contribution follows a different, double-peaked pattern. We note that the emitting region(s) responsible for the non-thermal emission are located somewhere in the neutron star magnetosphere, while the thermal emissions are coming, most probably, from close to the neutron star surface. Thus, the lack of phase correlation between the two thermal emissions is more surprising, and revealing, than the phase behavior of the higher energy photons.

While the cool black-body covers a sizable fraction of the neutron star surface, the hotter one comes from a minute fraction of it. Its inferred dimension (up to 60 m radius) is compatible with the expected extension of the polar cap for a spinning neutron star like Geminga (nominal value of the "dipole" polar cap radius, $R\sqrt{\frac{R\Omega}{c}}$ ~300 m, where R is the 10 km star radius and $\Omega$ is Geminga's angular velocity), especially considering possible geometrical effects[17]. A nearly aligned rotator viewed at high inclination could easily yield a geometrical reduction of ~10 between the nominal value and our apparent spot radius.

Thus, we propose the hot black-body emission to come from the polar cap(s) of the magnetized neutron star, heated by particles accelerated in the pulsar magnetosphere. The same process responsible for the copious gamma-ray emission of Geminga would thus also be responsible for the appearance of the hot spots on its surface[7]. Such a mechanism had been posited to exist for γ-ray pulsars [18,19].

The inferred bolometric luminosity of Geminga's polar cap(s) is 1.5 x 10$^{29}$ erg/sec. It is a value compatible with the predictions of [20] for polar cap heating based on inverse Compton scattering (ICS) pair fronts. Because heating from return curvature radiation would exceed our measured luminosity by at least two orders of magnitude[7,20], some mechanism must prevent it. Geminga is the only γ-ray pulsar close to the so called death-line, where pulsars no longer produce pairs by curvature radiation[20] and its age and evolution may help explain why the heating is diminished. We cannot determine how many hot-spots there are at one polar cap or whether hot spots occur at both caps. Three dimensional modelling, taking gravitational light bending into account, is needed to determine the hot spot distribution . The visibility of small hot spot(s) for the majority of the pulsar period suggests a nearly aligned rotator seen at high inclination.

The sinusoidal evolution of the cooler black-body favors a single extended polar region, warmer than the rest of the star because of higher thermal conductivity parallel to the magnetic field [21]. Such a geometry would yield a correlated behavior between the two black-body emissions. Instead, the lack of correlation in the phase evolution of the cool and hot black-body emissions from the surface of Geminga argues against such an interpretation, unless different beaming mechanisms are invoked for different emitting regions. Alternatively, the cool black-body could come from vast "continent" on the neutron star equatorial surface, where multipole magnetic fields could play a role[22]. Cold black-body radiation could also arise from surface x-rays reprocessed within a blanket formed above vast star regions by $e^+e^-$ pairs created on closed field lines[19]. The double-peaked structure of photons >2keV argues against the single-pole hypothesis. However, emitting regions higher up in the magnetosphere, at the rim of a single cone, could yield a two-peak non-thermal light curve[23]. Alternatively, the off-center dipole proposed by [7] could provide a viable solution. In a few years, by operating XMM-Newton in coordination with AGILE[24] and GLAST[25], it will be possible to exploit the physical link now seen between γ-rays, pulsar energetic particles, return current and polar hotspots to gain deeper insight into the complex phenomenology of Geminga.

26) The XMM-Newton data analysis is supported by the Italian Space Agency (ASI).

27) ADL acknowledges an ASI fellowship

**Figure captions**

Figure 1

Time-averaged spectral distribution of all the Geminga pn data. Fits were carried-out in the energy range 0.3-8 keV, where the instrument performances are best known. Background was subtracted after averaging over a suitable surrounding region. A combination of a single black-body and a power-law ($T_{bb}$= 44.5 ± 1 eV covering a surface with a 7 ±1 km radius; power-law photon index 1.90 ± 0.05; $N_H$ < 1.4 $10^{20}$ cm$^2$) yields a $\chi^2$= 1.35 over 73 d.o.f.. Adding a second black-body component to the spectral fit, we obtain:

$T_{bb1}$ = 43.8 ± 1 eV , radius of the emitting region 7.5 ±1 km;

$T_{bb2}$ =185 ± 20 eV, radius of the emitting region 33 ±10 m;

power-law photon index 1.7 ± 0.1 ; $N_H$ < 1.2 $10^{20}$ cm$^2$ , yielding a $\chi^2$ 1.21 over 71 dof. Using an F-test, such a fit improvement has a chance occurrence probability of 5 $10^{-3}$.

The corresponding observed flux is 1.1 $10^{-12}$ erg/cm$^2$ sec (0.3-8 keV) while the bolometric luminosities of the three components are : $L_{bb1}$ = 2.6 $10^{31}$ erg/sec ; $L_{bb2}$ = 1.5 $10^{29}$ erg/sec; $L_{pl\ (2\text{-}8\ keV)}$ = 7.7 $10^{29}$ erg/sec.

Figure 2

Background subtracted light curves, obtained, for three energy ranges, using a period of 0.2371012153 sec, extrapolated from the EGRET ephemeris [9]. Two phases are shown for clarity. The 10 phase intervals used for the time resolved spectroscopy are also indicated. Pulsed fractions have been computed as the ratio between the counts above the minimum of each light curve and the total number of counts detected in each energy interval.

Figure 3A

Ten color-coded phase resolved spectra of Geminga.

Figure 3B

Three-component fits of the 10 phase resolved spectra show the changing contribution of the cool and hot black-bodies and hard power-law. Note the strong variability (and disappearance) of the hot black-body tracing the hot spot(s) rotating on the surface of Geminga.

Figure 4

Phase evolution of the three components of Fig.3B. The variations of the apparent surfaces of the two black-body emitting regions as well as the non-thermal flux are shown.

Panel: 1 phase evolution of the radius (in km) of the emitting area of the 43 eV black-body.

Panel: 2 phase evolution of the radius (in m) of the emitting area of the 170 eV black-body.

Panel: 3 phase evolution of the power-law flux at 1 keV.

Figure S1

Phase-resolved spectra computed for the phase intervals shown in figure 2. The data points (crosses) are compared with a template reproducing the average spectrum of figure 1. Binning has been adjusted in order to have at least 80 counts per channel.

Figure S2

Differences, computed in unit of standard deviations, between the actual data and the average template for the 10 phase intervals. The biggest effect is seen in spectrum # 2, with several bins showing >10 σ deviations.

Videoclip   http://www.mi.iasf.cnr.it/~carapat/geminga/phase-resolved.DIVX.avi

The video shows variations of the raw Geminga EPIC spectrum for each phase interval. It then shows individual variations of the three components of our best fit. An inset shows the light curve in order to link phase spectral evolution with total source flux.

Fig 1

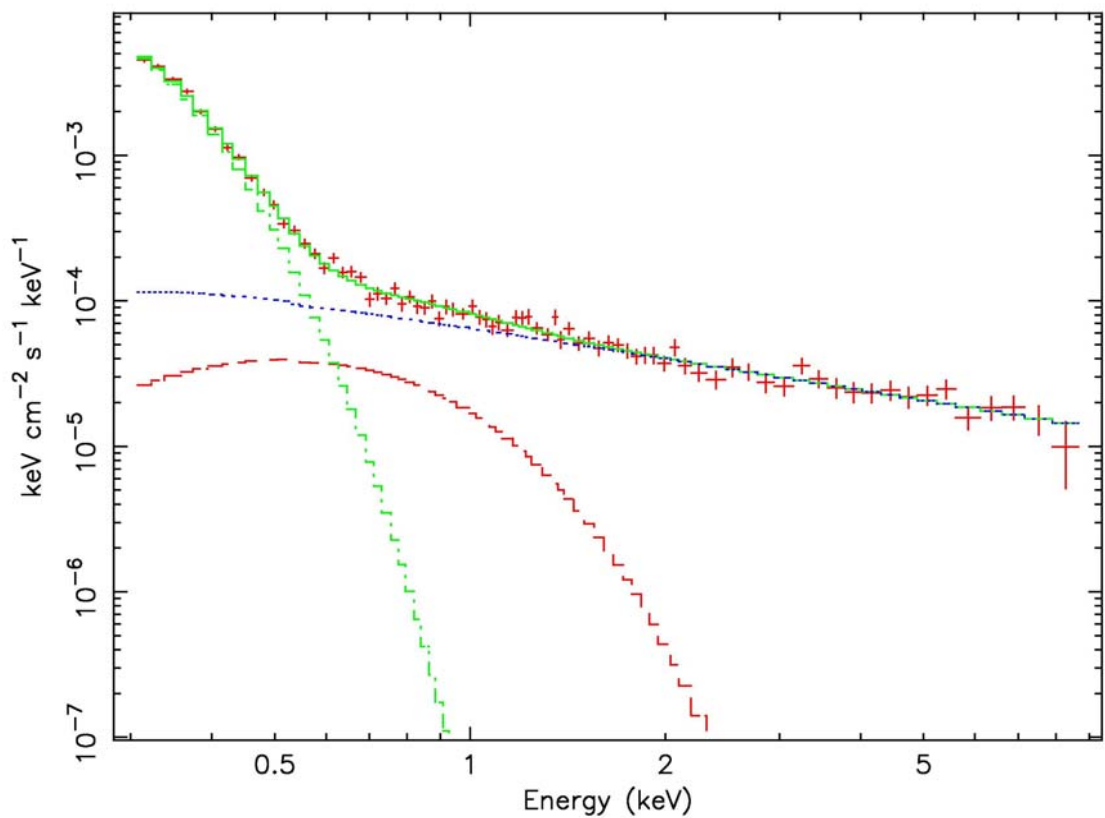

Fig 2

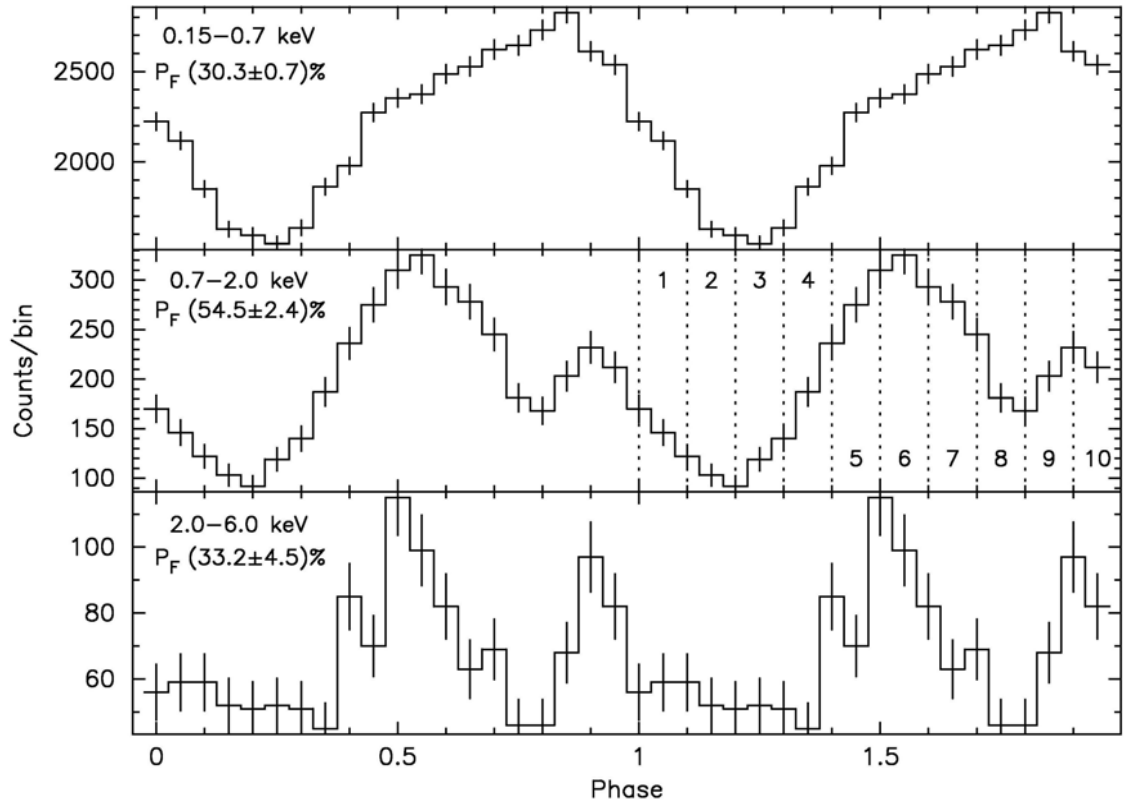

Fig 3a

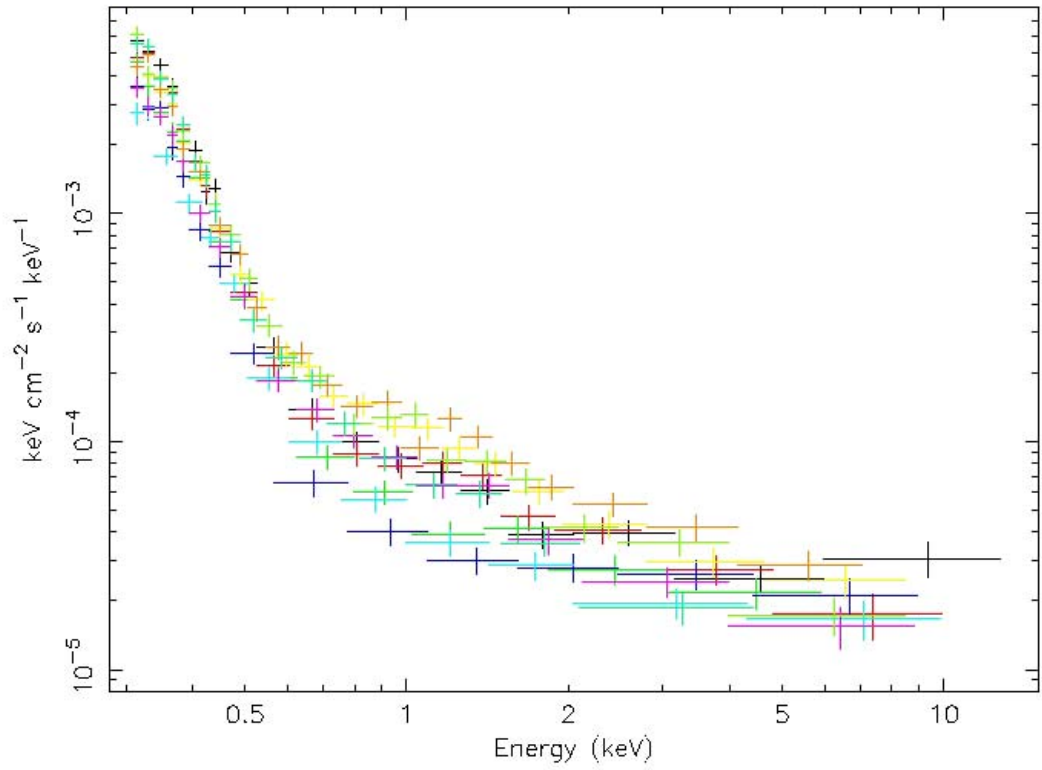

Fig3b

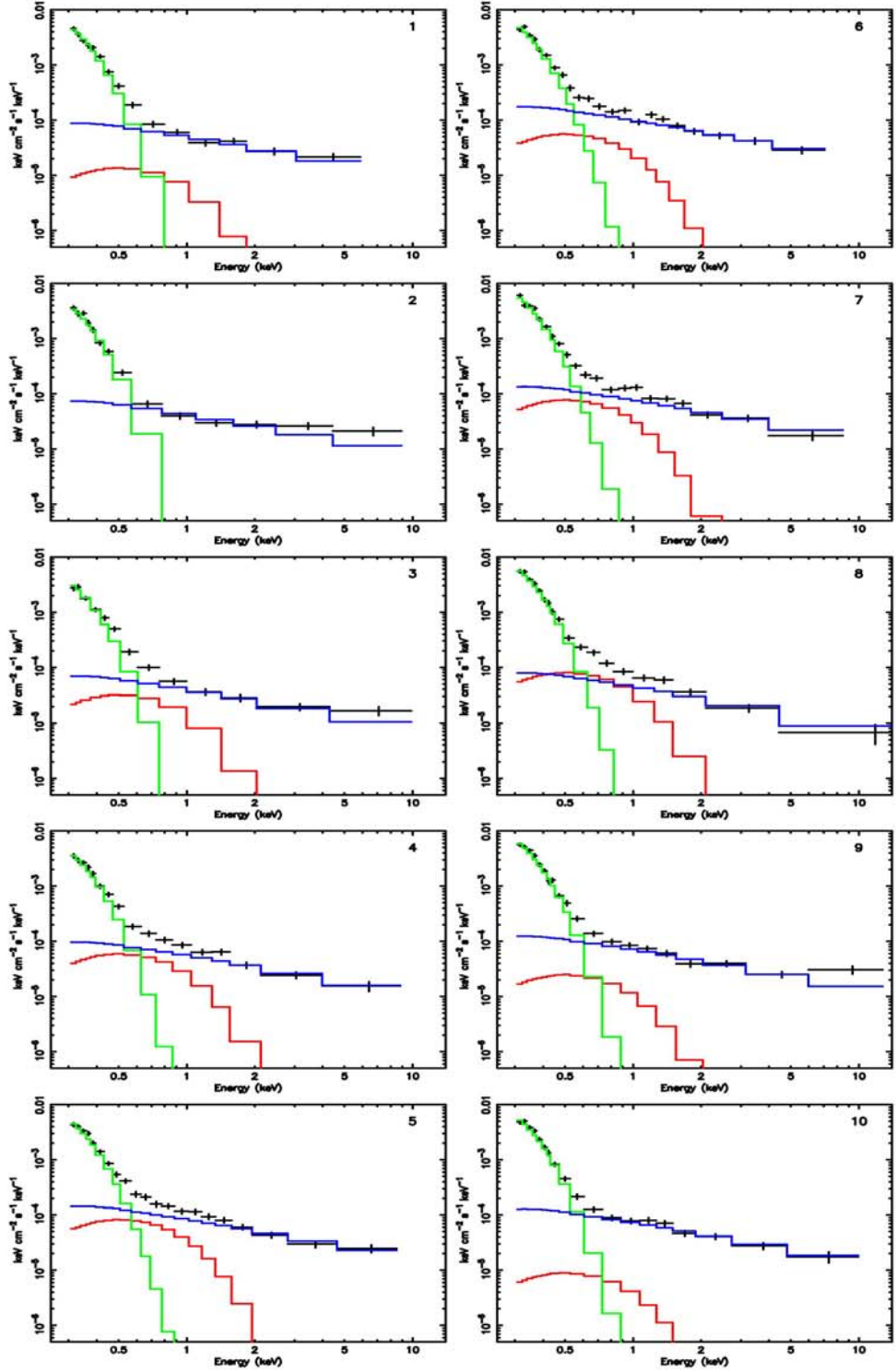

Fig 4

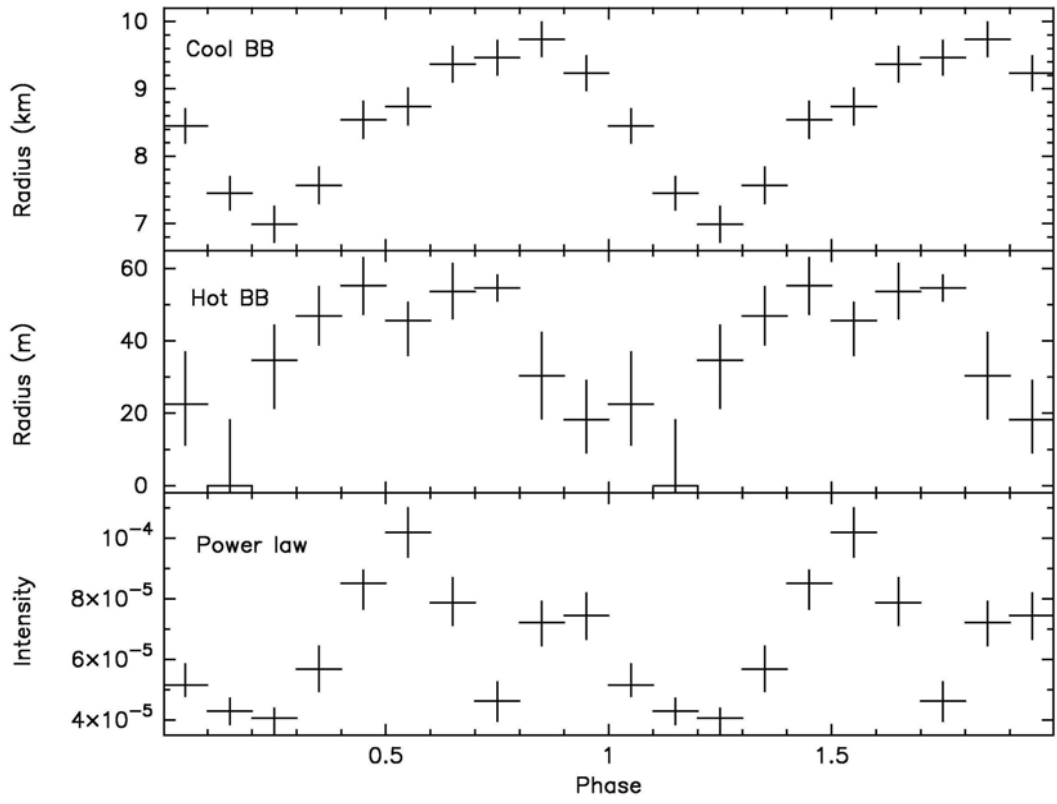

Fig S1

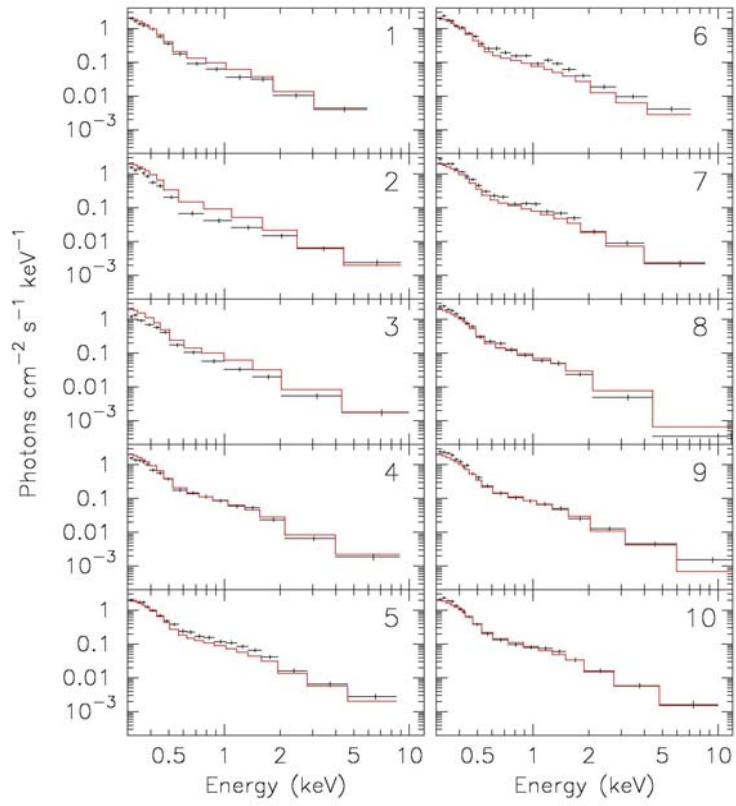

Fig S2

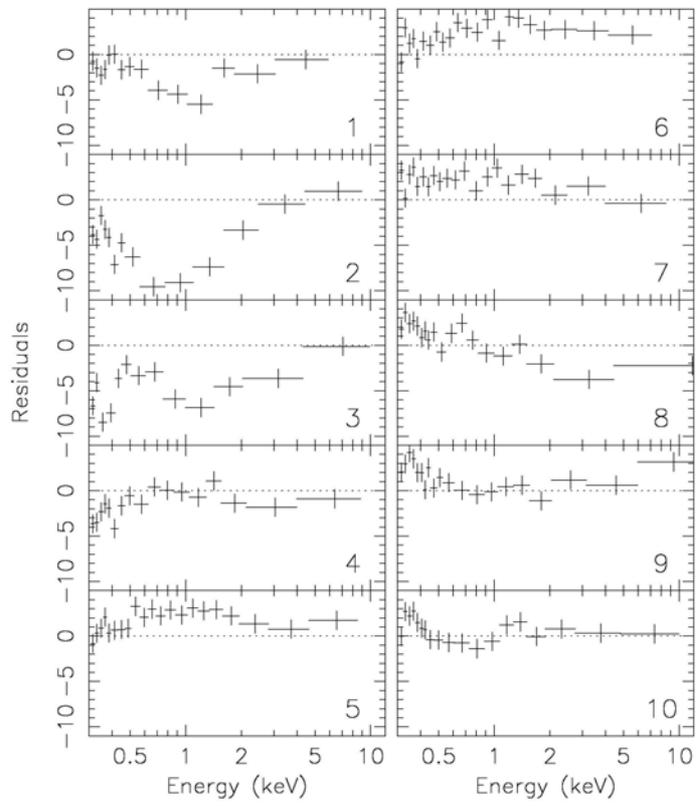